\documentstyle[aps,epsf]{revtex}
\begin{document} 

\draft

\title{Raman Sideband Cooling in presence of Multiple Decay Channels}
\author{G. Morigi, H. Baldauf, W. Lange, and H. Walther}
\address{Max Planck Institut f\"ur Quantenoptik, Hans-Kopfermannstrasse 1, D-85748 Garching, Germany}
\date{\today}
\maketitle

\begin{abstract}
We have investigated the efficiency of pulsed Raman sideband cooling in
the presence of multiple decay and excitation channels. By applying sum
rules we identify parameter regimes in which
multiple scattering of photons can be described by an effective wave
vector. Using this method we determine the rate of heating caused by
optical pumping inside and outside the Lamb-Dicke regime.
On this basis we discuss also the efficiency of a recently proposed scheme
for ground-state cooling outside the Lamb-Dicke regime [G. Morigi, J.I.
Cirac, M. Lewenstein, and P. Zoller, Europhys. Lett. {\bf 39}, 13
 (1997)].
\end{abstract}


\section{Introduction}

Laser-cooling \cite{Nobel} allows to cool ions and atoms to very low temperatures.
For this purpose, the full knowledge of the effects of the various physical parameters 
determining the cooling
process is very important. Among the various schemes, Raman sideband cooling has been 
demonstrated to be a very successful technique for preparing atoms in the ground state
of a harmonic potential \cite{RamanS}. This cooling method exploits two stable or metastable atomic 
internal levels, which we call $|g\rangle$ and $|e\rangle$, 
connected by dipole transitions to a common excited state $|r\rangle$. The transitions
are usually driven by alternating pulses.
A typical sequence alternates a coherent pulse, in which the atom is coherently transferred from 
$|g\rangle$ to $|e\rangle$ via a properly designed Raman pulse, with a re-pumping pulse,
in which the atom is incoherently re-scattered to $|g\rangle$ by means of a laser resonant with
$|e\rangle\to|r\rangle$. A change of the motional state during the repumping is a process of higher
order in the ratio $\omega_R/\nu$ of the recoil frequency $\omega_R=\hbar k^2/2m$ and the 
trap frequency $\nu$, with $m$ being the mass of the atom and $k$ the wave vector 
of the one-photon transition. In the Lamb-Dicke regime, where $\omega_R \ll\nu$,
the probability for a change of the motional state is negligible and therefore, 
on the average, the system is cooled at a rate
of one phonon of energy $\hbar\nu$ per cooling cycle. Since there is a
finite probability for the atom to be returned to the state $|e\rangle$ instead of being repumped,
a number of incoherent scattering events may be required
before the atom is finally scattered into $|g\rangle$,
which significantly increases the motional energy at the end of the optical pumping,
reducing the cooling efficiency.
Furthermore, since two and three level schemes are realized using Zeeman or hyperfine substates,
decays from $|r\rangle$ into other electronic substates can occur, leading to additional heating.\\
\indent In this work we quantify the effect of a finite branching ratio  
in pulsed Raman sideband cooling by calculating the average shift and
diffusion of the vibrational energy distribution
at the end of an incoherent pumping pulse.
It should be pointed out that theoretical studies on laser-cooling for multilevel ions exist, 
which systematically include the branching ratio in their treatments
\cite{Marzoli,Ignacio,3lev}. Those studies have focussed on the Lamb-Dicke regime and on certain
cooling schemes.
Here, we single out the effect of the branching ratio on cooling for an arbitrary ratio
$\omega_R/\nu$ by applying sum rules.
Hence, we infer the cooling efficiency in the Lamb-Dicke regime and we discuss the result outside 
the Lamb-Dicke regime in connection with the proposal in \cite{outLDL}.
In particular, we show that in some parameter ranges the average effect of the
multiple photon scattering
can be described with an effective wave vector $k_{\rm eff}$ for 
the ``effective'' two-level transition $|e\rangle\to |g\rangle$ \cite{Marzoli}.\\
\indent This article is organized as follows. In Section 2 we introduce
the model for the evolution of a trapped ion during the repumping pulse 
in a Raman transition, and we evaluate
the average shift and variance of the ion energy at the end of the pulse. 
In Section 3 we extend our analysis to cases where the channels of decay are
multiple. In Section 4 we draw some conclusions, and in the Appendix 
we report the details of our calculations.

\section{Model}

\begin{figure}
\begin{center}
\epsfxsize=0.3\textwidth
\epsffile{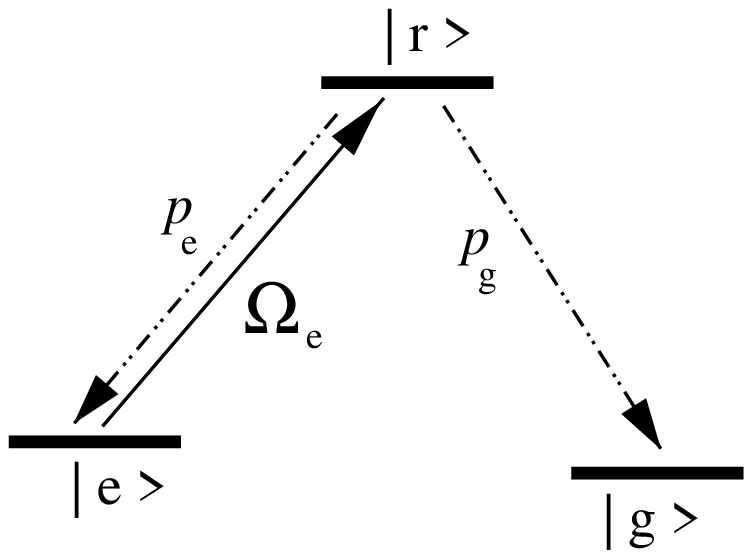}
\begin{caption}
{
Level scheme.
}
\label{fig:scheme0}
\end{caption}
\end{center}
\end{figure}

We consider a three level atom as in Fig. \ref{fig:scheme0}, 
whose internal levels are a ground state $|\text{g}\rangle$,
stable or metastable state $|\text{e}\rangle$ and excited state 
$|\text{r}\rangle$ of
radiative width $\gamma$; $|\text{g}\rangle\to|\text{r}\rangle$, 
$|\text{e}\rangle\to|\text{r}\rangle$ are
dipole transitions, with respective
probabilities of decay $p_g$,
$p_e$, where $p_g+p_e=1$. A laser resonantly drives the transition 
$|\text{e}\rangle\to|\text{r}\rangle$ with Rabi frequency
$\Omega_e$. In the following 
we assume the wave vectors for both
transitions to be equal to $k$,
which is a good approximation if, e.g., $|\text{e}\rangle$ and 
$|\text{g}\rangle$ are hyperfine components of the ground state.
We study the ion motion in one-dimension.\\
\noindent The master equation for the atomic density matrix $\rho_3$ 
is written as ($\hbar=1$):

\begin{equation} 
\frac{\text{d}}{\text{d}t}\rho_3=-i\left[H_0+V,\rho_3\right]+L\rho_3,
\label{Master0}
\end{equation}
where $H_0$ has the form:

\begin{equation}
H_0=\delta |e\rangle\langle e|+\nu a^+a.
\end{equation}
\noindent Here, $\delta$ is the detuning of the laser on the $|e\rangle\to |r\rangle$
transition, which we take to be zero, and $\nu$ is the frequency 
of the harmonic oscillator which traps the ion along the $x$-direction,
with $a,a^+$ annihilation and creation operator, respectively. The interaction of the 
ion with
the laser light is described in the dipole approximation by the operator $V$:

\begin{equation}
V   =\frac{\Omega_e}{2}\left(A_e^{+}\text{e}^{ikx}+\text{h.c.}\right),
\label{coherent1}
\end{equation}
with $A^+_j=|r\rangle\langle j|$ (with $j=e,g$) dipole raising operator,
$A_j^{-}$ its adjoint, and $x$ the position of the atom. 
In writing (\ref{coherent1}), (\ref{Master0}) 
we have applied the Rotating Wave Approximation 
and we have moved to the inertial frame rotating at the laser frequency.
Finally, the relaxation super--operator has the form

\begin{eqnarray}
L\rho_3&=&
-\frac{\gamma}{2}\left(|\text{r}\rangle\langle \text{r}|\rho_3 + 
\rho_3 |\text{r}\rangle\langle \text{r}| \right)\nonumber\\
       &+&\sum_{j=g,e}p_j\gamma\int_{-1}^{1}\text{d}uN(u)A_j^{-}\text{e}^{-ikux}
\rho_3\text{e}^{ikux}A_j^{+}\nonumber,
\end{eqnarray}
\noindent where $N(u)$ is the dipole pattern of the spontaneous emission, 
which we take $N(u)=3/8( 1 + u^2 )$.\\
In the limit $\Omega_e\ll \gamma$ we can eliminate
the excited state $|r\rangle$ in second order perturbation theory \cite{Gardiner}, 
and reduce the three-level scheme to a two level one, with excited state 
$|\text{e}\rangle$ and linewidth $\gamma_e=\Omega_e^2/\gamma$ \cite{Marzoli}. 
In the limit $\gamma\gg\nu$ the master equation for the 
density matrix $\rho$, projection of $\rho_3$ on the subspace  
$\{|e\rangle,|g\rangle\}$, can be rewritten as
\cite{MasterDumm}:

\begin{equation} 
\frac{\text{d}}{\text{d}t}\rho=-i\left[H_{\rm eff}\rho
-\rho H_{\rm eff}^{+}\right]+\gamma_e\left[J_e\rho
+J_g\rho\right],
\label{Master1}
\end{equation}
with $H_{\rm eff}$ effective Hamiltonian

\begin{equation}
\label{Heff}
H_{\rm eff}=H_0-i\frac{\gamma_e}{2}
|\text{e}\rangle\langle \text{e}|,
\end{equation}
and with $J_e\rho$, $J_g\rho$ jump operators, defined as:

\begin{equation}
\label{Jump}
J_j\rho=p_j\left(\sigma_{je}\left[J^{\prime}\rho\right]\sigma_{ej}\right) \mbox{~~~with~~~}j=g,e,
\end{equation}
where $\sigma_{ij}=|i\rangle\langle j|$ and where 

\begin{equation}
J^{\prime}\rho=\int_{-1}^{1}\text{d}uN(u)
\text{e}^{-ik(1+u)x}\rho\text{e}^{ik(1+u)x}.
\end{equation}

\noindent The solution of Eq. (\ref{Master1}) can be written as follows \cite{MasterDumm}:

\begin{eqnarray}
\label{Sol}
\rho(t)&=    &S(t)\rho(0)+\gamma_e\int_0^t\text{d}t_1S(t-t_1)JS(t_1)\rho(0)+...\\
       &+    &\gamma_e^n\int_0^t\text{d}t_1\int_0^{t_1}\text{d}t_2...\int_0^{t_{n-1}}
\text{d}t_n  S(t-t_1)JS(t_1-t_2)J_e...J_eS(t_{n})\rho(0)+...~,
\nonumber
\end{eqnarray}
with $J=J_e+J_g$, and $S(t)$ is the propagator for the effective Hamiltonian:

\begin{equation}
S(t)\rho(0)=\text{e}^{-iH_{\rm eff}t}
\rho(0)\text{e}^{iH_{\rm eff}^+t}.
\end{equation}
In Eq. (\ref{Sol}) the successive contributions to the multiple scattering event are
singled out: The first term on the RHS corresponds to the case in which
at time $t$ no spontaneous decay has occurred. The second term describes a single 
scattering event, and the $n$-th term 
$n-1$ scattering events. The trace of each term corresponds to the probability 
associated with each event,
and we can thus interpret Eq. (\ref{Sol}) as
the sum over all the possible paths of the scattering event weighted by their respective
probabilities. At $t\to\infty$, $\rho(t)\to \rho_S$ , the atom is in 
$|g\rangle$ and $\langle e|\rho_S|e\rangle=0$. For a pulse of duration
$t\gg 1/\gamma_e$ we can replace $t$ by $\infty$ in the
integrals of Eq. (\ref{Sol}) and assume that the atom has been scattered into 
$|g\rangle$ at the end of the pulse. Now, each term on the RHS of Eq. (\ref{Sol}) corresponds to the path
associated with a certain number of scattering events into 
$|e\rangle$ before the atom is finally scattered into $|g\rangle$.
Through (\ref{Sol}) we can evaluate 
the shift and the variance of the energy distribution at the end of the repumping pulse,
which are defined as:

\begin{eqnarray}
\label{shift0}
\Delta E&=&\mbox{Tr}\{\left(H_{\rm mec}-E_0\right)\rho_S\},\\
\sigma_E&=&
\sqrt{\mbox{Tr}\{\left(H_{\rm mec}-E_0-\Delta E\right)^2\rho_S\}},
\label{diff0}
\end{eqnarray}
where $H_{\rm mec}=\nu a^+a$ and 
$E_0$ is the initial motional energy of the atom.

\subsection{Evaluation of the average shift and diffusion}

For simplifying the form of the discussion presented below, 
we rewrite the operator $J^{\prime}$ as follows:

\begin{equation}
J^{\prime}\rho=\tilde{J}\rho+\hat{J}\rho,
\end{equation}
where $\tilde{J}$, $\hat{J}$ are defined as:

\begin{eqnarray}
\label{tilde}
&\tilde{J}\rho=\sum_l |l\rangle\langle l|\left[J\rho\right]|l\rangle\langle l|,&\\
&\hat{J}\rho=\sum_l\sum_{l_1,l_1\neq l} 
|l\rangle\langle l|\left[J\rho\right]|l_1\rangle\langle l_1|,&
\end{eqnarray}
and where  $\{|l\rangle\}$ is the basis of 
eigenstates of the harmonic oscillator.
For $\rho(0)=|e\rangle\langle e|\otimes \mu(0)$, with $\mu(0)$ initial
distribution over the motional states, and 
according to Eq. (\ref{Sol}) the steady state
distribution has the form:

\begin{equation}
\label{Sol1}
\mu_{\infty}
=\sum_{m=1}^{\infty}p_gp_e^{m-1}\tilde{J}^m\mu(0)+F(\tilde{J},\hat{J})\mu(0),
\end{equation}
where $\mu_{\infty}=\langle g|\rho_S|g\rangle$ is the final
distribution over the motional states.
The first term in the RHS of (\ref{Sol1}) is
the sum over all paths from $|e\rangle$ into $|g\rangle$, where 
after each jump the density operator is diagonal in 
the basis $\{|l\rangle\}$, whereas the second term contains all other paths. 
These latter terms can be neglected \cite{Secular}, and 
for $\mu(0)=|n\rangle\langle n|$ 
the following relation holds:

\begin{equation}
\langle s|\mu_{\infty}|s\rangle
\approx \langle s|\left[\sum_{m=1}^{\infty}
\tilde{J}^m |n\rangle\langle n| \right]|s\rangle=D_n(s).
\label{Distri}
\end{equation}
Here, $D_n(s)$ is the probability for the
atom to be found in the state
$|g,s\rangle$ at $t\to\infty$, 
given the initial state $|e,n\rangle$ at $t=0$. 
Using the 
explicit form (\ref{tilde}) of $\tilde{J}$ in (\ref{Distri}), 
$D_n(s)$ has the form:

\begin{eqnarray}
\label{explicit}
D_n(s)
&=    &  p_g\sum_{m=0}^{\infty}p_e^{m-1}
\sum_{k=0}^{\infty}
\cdot \int_{-1}^{1}\text{d}u_1...\int_{-1}^{1}\text{d}u_m
N(u_1)...N(u_m)\\
&\cdot& \sum_{k_1=0}^{\infty}...\sum_{k_{m-1}=0}^{\infty}
|\langle s|\text{e}^{i ka_0(1+u_1)(a^{\dagger}+a)}|k_1\rangle|^2...
|\langle k_{m-1}|\text{e}^{ika_0(1+u_m)(a^{\dagger}+a)}|n\rangle|^2,
\nonumber
\end{eqnarray}

\noindent where we have used the relation $x=a_0(a^++a)$, with 
$a_0=\sqrt{1/2m\nu}$ size of the ground state of the harmonic oscillator. 
Substituting (\ref{explicit}) into
Eqs. (\ref{shift0}), (\ref{diff0}), and applying the commutation properties 
of $a,a^+$ [see the Appendix], we find:

\begin{eqnarray}
\langle \Delta E\rangle&=&\nu \eta^2\frac{7}{5}\frac{1}{1-p_e},
\label{center1}\\ 
\sigma_E^2
&=&\nu^2\left[\eta^2 \frac{7}{5}(2n+1)\frac{1}{1-p_e} + 
\left(\frac{7\eta^2}{5}\right)^2
\frac{58}{49}\frac{p_e}{(1-p_e)^2}\right],
\label{width1} 
\end{eqnarray}
where $\eta=ka_0=\sqrt{\omega_R/\nu}$ is the Lamb-Dicke parameter. 

\subsection{Discussion}

Equation (\ref{center1}) represents the average shift to the vibrational 
energy at the end of the
repumping pulse. For $p_e=0$ it corresponds to the average
recoil energy $\omega_R'$ associated with one incoherent
Raman scattering into $|g\rangle$. In this case, the second term in the 
RHS of Eq. (\ref{width1}) vanishes, and Eqs. (\ref{center1}), (\ref{width1}) 
describe the scattering of one photon
of wave vector $k'=\sqrt{7/5}k$ on the effective two-level transition 
$|e\rangle\to|g\rangle$. Similarly for $p_e>0$ an effective wave vector $k_{\rm eff}$
can be defined for the incoherent scattering on the two-level transition $|e\rangle\to|g\rangle$,
which has the form

\begin{equation}
k_{\rm eff}=\frac{k'}{\sqrt{1-p_e}}=\sqrt{\frac{7}{5}}\frac{k}{\sqrt{1-p_e}}.
\label{eta_eff}
\end{equation}  
Thus, $ k_{\rm eff}$ describes the average mechanical effect on the ion 
resulting from 
the multiple scattering 
of photons during the repumping pulse in a Raman transition with
branching ratio $(1-p_e)/p_e$:
This description is valid in the limit in which
we may neglect the second term in the RHS of (\ref{width1}), {\it i.e.}
for $p_e$ and/or $\eta$ sufficiently small. 
In Fig. 2 the first term of RHS of Eq. (\ref{width1}) is compared
with the complete expression for $n=0$, for different values
of the Lamb-Dicke parameter and as a function of $p_e$. Here, we see that $k_{\rm eff}$
characterizes the scattering process for almost any branching ratio in the Lamb-Dicke regime, 
whereas for $\eta=0.6$ an appreciable difference is already visible at $p_e=0.2$.\\
\noindent From (\ref{eta_eff}) we can define the
effective Lamb-Dicke parameter $\eta_{\rm eff}=k_{\rm eff}a_0$ describing 
an incoherent scattering into the state $|g\rangle$. This parameter
provides an immediate estimate of the effect of the branching ratio on cooling. For $\eta\sqrt{n}\ll 1$, 
if $\eta_{\rm eff}\sqrt{n}\ll 1$ the system is still in the Lamb-Dicke regime once it has been finally 
scattered into $|g\rangle$. Furthermore, 
the coarse-grained dynamics of the system can be described by a rate equation
for the motional states $|n\rangle$ projected onto $|g\rangle$, 
where the rate of cooling (heating) is the real part of the sum of two terms:
one corresponding to the component of the fluctuation spectrum of 
the dipole force at frequency $\nu$ ($-\nu$), the other to the diffusion coefficient 
due to spontaneous emission from the excited state \cite{Marzoli,Ignacio}. This latter term is 
proportional to the squared
Lamb-Dicke parameter for the incoherent scattering, and thus in our case
to $\eta_{\rm eff}^2$. 
From the well-known solution of the rate equation \cite{Stenholm}, 
the diffusion term affects the steady state
average vibrational number $\langle n\rangle$, which is proportional to the diffusion coefficient.\\
Outside the Lamb-Dicke regime, when $\omega_R$ is comparable to, or larger than, $\nu$, there
are no estabilished ground-state laser-cooling techniques for trapped atoms. 
Here, we discuss our result
in connection to the proposal in \cite{outLDL}. There, a cooling scheme similar to
Raman sideband cooling has been presented, where pulses which pump the atoms to the
ground state alternate with pulses confining the atoms to a limited region of 
motional energy. 
These confinement pulses have two-photon detuning $\delta_c$ to the red of the two-photon
resonance frequency, where $\delta_c \approx \omega_R'$. Then,
the presence of a branching ratio must be taken into account by choosing $\delta_c\approx \Delta E$.
In this regime, pulses which efficiently counteract the average kick $\Delta E$ can be designed,
provided that the following condition is fulfilled:

\begin{equation}
\label{Valid}
k_{x}^{\rm coh}\ge k_{\rm eff},
\end{equation}
where $k_{x}^{\rm coh}$ is the projection on $x$ of the two-photon wave vector of
the coherent pulse. For two counterpropagating beams parallel to $x$, 
$k_{x}^{\rm coh}=2|k|$
and (\ref{Valid}) is fulfilled for $p_e\le 13/20$, i.e. up to branching ratios $p_e/p_g\approx 2$. 
Finally, outside the Lamb-Dicke regime
the second term in the RHS of Eq. (\ref{width1}) cannot be neglected.
Hence, the diffusion is larger, and the efficiency of cooling may
decrease dramatically as $p_e$ increases.

\section{Extension to multi-level schemes}

In the following, we show that the average heating associated with the
repumping pulse in multilevel-schemes can be described in the same way as
discussed in the previous sections. \\
Let us consider the level-scheme of Fig. 3(a), where we have added to the scheme of Fig. 1
a further channel of decay from $|r\rangle$ into the stable or metastable state $|1\rangle$,
with probability of decay $p_1$ such that $p_1+p_e'+p_g'=1$, where $p_e'$, $p_g'$
are the probability of decay onto $|e\rangle,|g\rangle$, respectively. A laser
resonantly drives the transition
$|1\rangle\to |r\rangle$ with Rabi frequency $\Omega_1$.
For $\Omega_e,\Omega_1\ll \gamma$ the state $|r\rangle$ can be
adiabatically eliminated from the equations of motion. 
In this limit the Master Equation aquires the form 

\begin{eqnarray}  
\label{MasterMany}
&&\frac{\text{d}}{\text{d}t}\rho=-i\left[H_{\rm eff}\rho
-\rho H_{\rm eff}^{+}\right]\\
&&+p_e'\gamma^{\prime}J_e\rho
+p_1\gamma^{\prime}J_1\rho
+p_g'\gamma^{\prime}J_g\rho,\nonumber
\end{eqnarray}

\noindent where $\gamma^{\prime}=\gamma_e+\gamma_1$, with $\gamma_j=\Omega_j^2/\gamma$.
The effective Hamiltonian is now:

\begin{equation}
\label{HeffMany}
H_{\rm eff}=H_0-i\frac{\gamma_e}{2}
|\text{e}\rangle\langle \text{e}|-i\frac{\gamma_1}{2}
|\text{1}\rangle\langle \text{1}|,
\end{equation}
and the jump operators have the form:

\begin{equation}
\label{JumpMany}
J_i\rho=p_i\sum_{j=1,e}\sigma_{ij}\left(\tilde{J}+\hat{J}\right)\sigma_{ji},
\end{equation}
with $i=1,e,g$.
The solution at $t\to\infty$ can be written as:

\begin{equation}
\label{SolMany}
\mu_{\infty}=\sum_{m=1}^{\infty}p_g'(p_e'+p_1)^{m-1}\tilde{J}^m\mu(0).
\end{equation}
Hence, the shift and variance have the form evaluated in Eqs. (\ref{center1}), (\ref{width1})
where now the probability $p_g$, $p_e$ are defined as
$p_e=p_e'+p_1$, $p_g=p_g'$ ($p_g+p_e=1$).
In a similar way we have evaluated these quantities for schemes like the one shown in fig. 3(b), 
where a second excited state $|2\rangle$ is coupled to
$|e\rangle$ via the same recycling laser tuned on the transition
$|1\rangle\to|r\rangle$. For simplifying the treatment, we assume that a fourth laser 
resonantly drives the transition $|1\rangle\to |2\rangle$
with Rabi frequency $\Omega$ (grey arrow in fig. 3(b)). 
Thus, for low saturation Eq. (\ref{MasterMany}) describes 
the dynamics, where now $\gamma_i=\gamma_i^{(r)}+\gamma_i^{(2)}$ ($i=e,1$), with
$\gamma_i^{(j)}$ being the rate of scattering through the excited state $|j\rangle$ ($j=r,2$).
Assuming that $\Omega$ is such that $\gamma_e^{(r)}/\gamma_e^{(2)}=\gamma_1^{(r)}/\gamma_1^{(2)}=a$,
the solution in Eqs. (\ref{center1}), (\ref{width1}) applies to this
case too, where now  $p_e$ is defined as:

\begin{equation}
p_e=(p_e'+p_1)\frac{a}{1+a}+\frac{1}{1+a},
\label{result_prob}
\end{equation}
and the probability $p_g$ of decaying into $|g\rangle$ is $p_g=1-p_e$.\\
The result (\ref{result_prob}) shows that the total heating is minimum
for $a\gg 1$, which can be obtained by
choosing properly the laser intensity of the repumping lasers, or simply by removing
degeneracies in the Zeeman multiplet, 
for example with the help of a magnetic field.

\section{Conclusions}

We have studied the motional heating associated with a finite 
branching ratio and in the presence of multiple decay and excitation channels
at the end of a repumping pulse in Raman sideband cooling.  
The first and second moments of the final energy distribution 
has been evaluated analytically, and the effect of
the branching ratio has been singled out. We have shown that in a certain range of 
parameters the diffusion can be 
described with an effective wave vector $k_{\rm eff}$, corresponding to an 
effective Lamb--Dicke parameter $\eta_{\rm eff}$ for the incoherent scattering on the
two-level transition $|e\rangle\to |g\rangle$. Finally, on the basis of this result 
we have discussed the efficiency of Raman sideband cooling and of a recent proposal
of ground-state cooling outside the Lamb-Dicke regime \cite{outLDL}.\\
Analogous sum rules and considerations can be applied to Raman cooling for 
free atoms \cite{Raman}. In that case the calculations are much simpler, since the total
momentum of radiation and atom is a conserved quantity in the scattering event. \\
In general, these results can be applied to cooling schemes in multilevel atoms.

\section{Ackwoledgements}
The authors acknowledge many stimulating discussions with S. K\"ohler and V. Ludsteck.
G.M. thanks J.I. Cirac, J. Eschner and P. Lambropoulos for many stimulating discussions. 
This work is supported in parts by the European Commission  within the TMR-networks 
ERB-FMRX-CT96-0087 and ERB-FMRX-CT96-0077.


\section{Appendix}

\noindent Using (\ref{Distri}), we rewrite 
(\ref{shift0}) and (\ref{diff0}) as:

\begin{eqnarray}
\label{Adi:center} 
\langle \Delta E\rangle&=&\nu\sum_{m=0}^{\infty}B_m^n,\\
\label{Adi:width} 
\sigma_E^2		 &=&\nu^2\sum_{m=0}^{\infty}A_m^n,
\end{eqnarray}
where we have introduced the quantities

\begin{eqnarray}
\label{Am}
A_m^n&=&
\sum_{s=0}^{\infty}(s-n-\Delta E/\nu)^2\langle s|\left[\tilde{J}^{m}|n\rangle\langle n|\right]|s\rangle,
\\
\label{Bm}
B_m^n&=&
\sum_{s=0}^{\infty}(s-n)\langle s|\left[\tilde{J}^{m}|n\rangle\langle n|\right]|s\rangle.
\end{eqnarray}
Using Eq. (\ref{explicit}), Eq. (\ref{Bm}) is rewritten as:

\begin{eqnarray}
B_m^n
&=    &p_gp_e^{m-1}\sum_{k_1=0}^{\infty}
(k_1-n)\int_{-1}^{1}
\text{d}u_1...\int_{-1}^{1}\text{d}u_m
N(u_1)...N(u_m)\nonumber\\
&\cdot&\sum_{k_2=0}^{\infty}...\sum_{k_{m}=0}^{\infty}
|\langle k_1|\text{e}^{i\eta(1+u_1)(a^{\dagger}+a)}|k_2\rangle|^2...
...|\langle k_{m}|\text{e}^{i\eta(1+u_m)(a^{\dagger}+a)}|n\rangle|^2.
\label{B}
\end{eqnarray}
The sum over $k_1$ can be contracted by observing that 
$k_1|k_1\rangle\langle k_1|=a^{\dagger}a|k_1\rangle\langle k_1|$. 
Then, using the commutation properties of the
bosonic operators $a,a^{\dagger}$ 
and the closure relation for the eigenstates of the
harmonic oscillator, Eq. (\ref{B}) takes the form:

\begin{eqnarray}
\label{Bbb}
B_m^n
&=    & p_gp_e^{m-1}\int_{-1}^{1}
\text{d}u_1...\int_{-1}^{1}\text{d}u_m
N(u_1)...N(u_m)\\ 
(-n
&+    &\eta^2(1+u_1)^2+
\sum_{k_2=0}^{\infty}...\sum_{k_{m-1}=0}^{\infty}
k_2
|\langle k_2|\text{e}^{i\eta(1+u_2)(a^{\dagger}+a)}|k_3\rangle|^2...
|\langle k_{m-1}|\text{e}^{i\eta(1+u_m)(a^{\dagger}+a)}|n\rangle|^2).
\nonumber
\end{eqnarray}
Repeating then the procedure shown in Eqs. (\ref{B}),(\ref{Bbb}) for each
index $k_i$, we finally obtain:

\begin{equation}
B_m^n= p_gp_e^{m-1}\frac{7}{5}\eta^2 m.
\label{Bnew}
\end{equation}
Analogously, $A_m^n$ has the form:

\begin{equation}
\label{Anew}
A_m^n
=    p_gp_e^{m-1}\left(\frac{7}{5}\eta^2 (2n+1) m+
\left(\eta^2\frac{7}{5}\right)^2\frac{29}{49} m(m-1)\right).
\end{equation}
Substituting now (\ref{Bnew}), (\ref{Anew}) into Eqs. (\ref{Adi:center}), (\ref{Adi:width})
and summing over $m$ we finally obtain Eqs. (\ref{center1}),
(\ref{width1}).

\begin{figure}
\begin{center}
\epsfxsize=0.3\textwidth
\epsfxsize=0.25\textheight
\epsffile{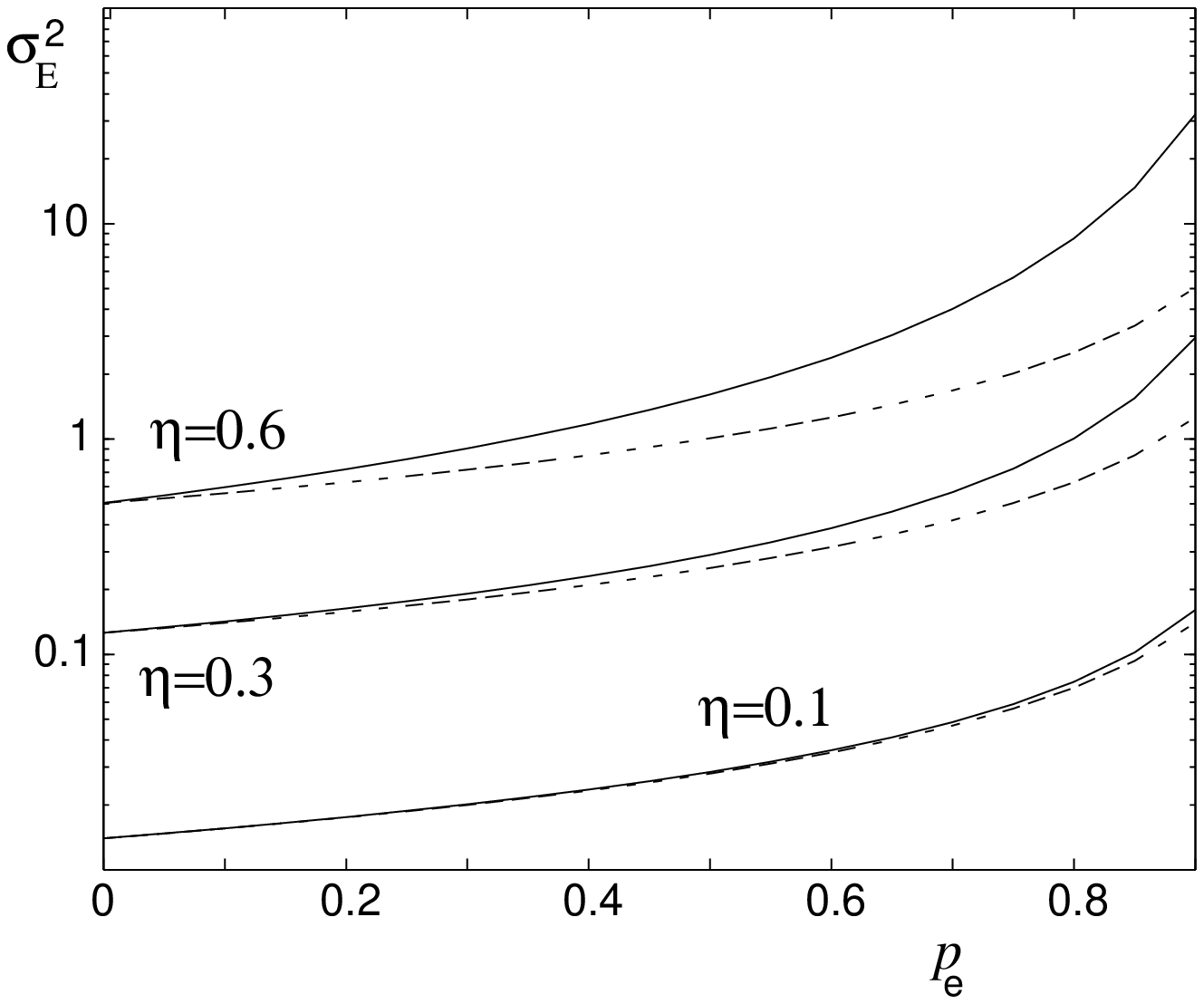}
\caption{
Plot of $\sigma_E^2$ (solid line), and of the first term on the RHS of Eq. (19) (dashed line)
as a function of $p_e$ for Lamb--Dicke parameter $\eta=0.1,0.3,0.6$ and for $n=0$.
}
\label{fig:branch2}
\end{center}
\end{figure}

\begin{figure}
\begin{center}
\epsfxsize=0.3\textwidth
\epsffile{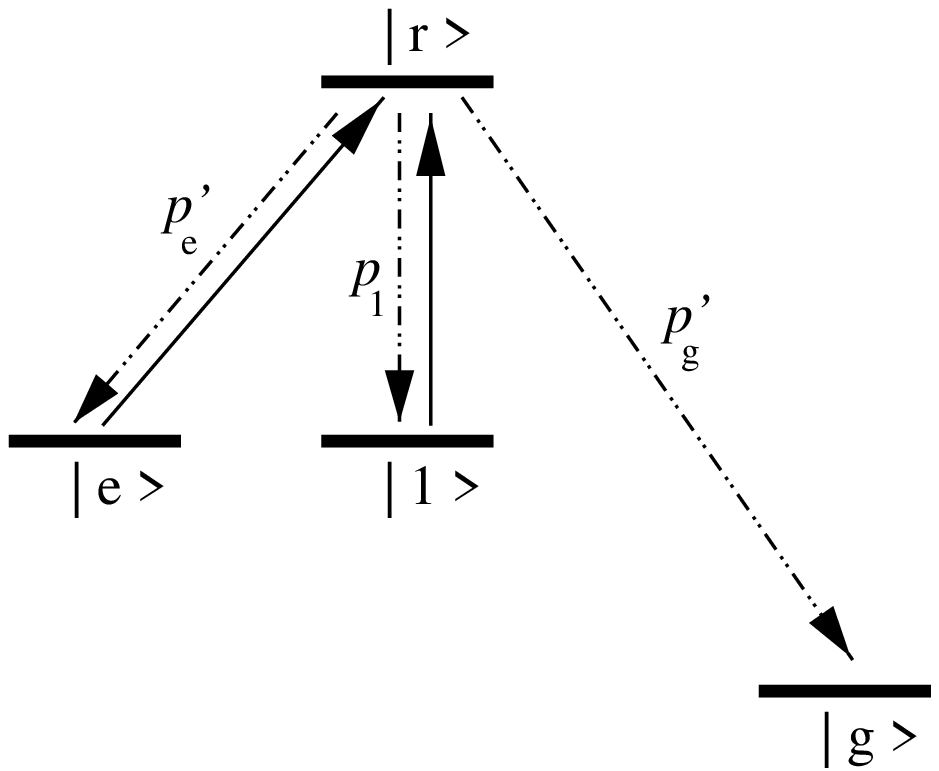}
\epsfxsize=0.3\textwidth
\epsffile{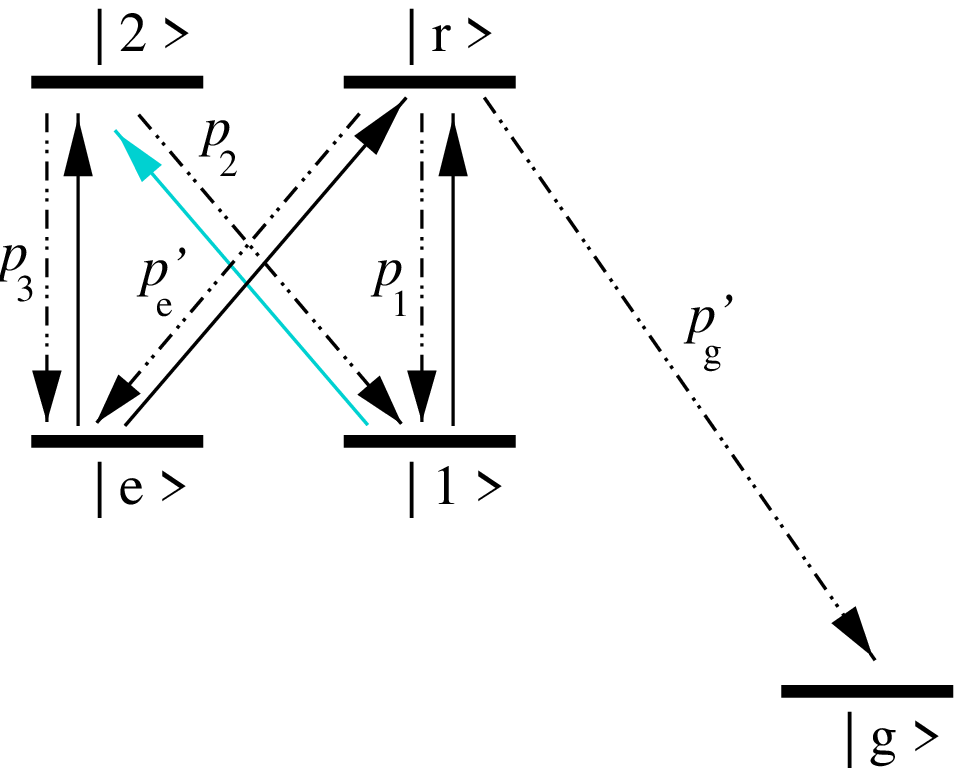}
\begin{caption}
{
(a) Level scheme with $|g\rangle$, $|e\rangle$, $|1\rangle$, stable or
metastable states, $|r\rangle$ excited state of radiative width
$\gamma$ and probability of decaying in the three ground states 
$p_g'$, $p_e'$ and $p_1$, respectively. Two lasers
couple $|e\rangle$ and $|1\rangle$ to $|r\rangle$; (b) Level scheme as
in (a) with the addition of the excited state $|2\rangle$ with decay probability
on $|1\rangle$, $|e\rangle$ equal to $p_2$, $p_3$, respectively, $p_2+p_3=1$. 
Two lasers couple $|e\rangle$ and $|1\rangle$ to $|2\rangle$. 
}
\end{caption}
\end{center}
\end{figure}


\begin{thebibliography}{99}
\bibitem{Secular}
It can be shown that the terms in Eq. (\ref{Sol}) containing at some
time $t'$ coherences between vibrational states
are of order $\gamma'/\nu$ with respect to the terms that contain 
the populations only. This condition alone would not be sufficient, as the number
of terms of $F$ corresponding to $n$ scattering events increases with $n$. However,
these term are oscillating functions of the intermediate
vibrational states, and thus their sum is much smaller than the first term of 
(\ref{Distri}).
\end{thebibliography}
\end{document}